\documentclass[preprint2]{aastex} %a double-column, single-spaced document:

\shorttitle{Enrichment of VMP Stars with {\boldmath $r$}-Process and {\boldmath $s$}-Process Elements}
\shortauthors{Wanajo et al.}

\begin{document}

\title{Enrichment of Very Metal-Poor Stars with Both {\boldmath
$r$}-Process and {\boldmath $s$}-Process Elements from {\boldmath $8-10
M_\odot$} Stars}

\author{Shinya Wanajo\altaffilmark{1}, Ken'ichi Nomoto\altaffilmark{2},
        Nobuyuki Iwamoto\altaffilmark{3}, Yuhri Ishimaru\altaffilmark{4},
        and Timothy C. Beers\altaffilmark{5}
} 

\altaffiltext{1}{Research Center for the Early Universe,
   Graduate School of Science, University of Tokyo,
   Bunkyo-ku, Tokyo 113-8654, Japan;
   wanajo@resceu.s.u-tokyo.ac.jp}

\altaffiltext{2}{Department of Astronomy, School of Science,
   University of Tokyo, Bunkyo-ku, Tokyo, 113-0033, Japan;
   nomoto@astron.s.u-tokyo.ac.jp}

\altaffiltext{3}{Nuclear Data Center,
   Japan Atomic Energy Research Institute, Ibaraki 319-1195, Japan;
   niwamoto@ndc.tokai.jaeri.go.jp}

\altaffiltext{4}{Academic Support Center, Kogakuin University,
   Hachioji, Tokyo 192-0015, Japan;
   kt13121@ns.kogakuin.ac.jp}

\altaffiltext{5}{Department of Physics \& Astronomy and JINA: Joint Institute
for Nuclear Astrophysics, Michigan State University,
E. Lansing, MI 48824, USA; beers@pa.msu.edu}

\begin{abstract}

Recent spectroscopic studies have revealed the presence of numerous
carbon-enhanced, metal-poor stars with [Fe/H] $< -2.0$ that exhibit
strong enhancements of $s$-process elements. These stars are believed to
be the result of a binary mass-transfer episode from a former asymptotic
giant-branch (AGB) companion that underwent $s$-process nucleosynthesis. 
However, several such stars exhibit significantly lower Ba/Eu ratios
than solar $s$-process abundances. This might be explained if there were
an additional contribution from the $r$-process, thereby diluting the
Ba/Eu ratio by extra production of Eu. We propose a model in which the
double enhancements of $r$-process and $s$-process elements originate
from a former $8-10 M_\odot$ companion in a wide binary system, which
may undergo $s$-processing during an AGB phase, followed by
$r$-processing during its subsequent supernova explosion. The mass of Eu
(as representative of $r$-process elements) captured by the secondary
through the wind from the supernova is estimated, which is assumed to be
proportional to the geometric fraction of the secondary (low-mass,
main-sequence) star with respect to the primary (exploding) star. We
find that the estimated mass is in good agreement with a constraint on
the Eu yield per supernova event obtained from a Galactic chemical
evolution study, when the initial orbital separation is taken to be
$\sim 1$ year. If one assumes an orbital period on the order of five
years, the efficiency of wind pollution from the supernova must be
enhanced by a factor of $\sim 10$. This may, in fact, be realized if the
expansion velocity of the supernova's innermost ejecta, in which the
$r$-process has taken place, is significantly slow, resulting in an
enhancement of accretion efficiency by gravitational focusing.

\end{abstract}

\keywords{
%--- 
nuclear reactions, nucleosynthesis, abundances
--- stars: abundances
--- stars: Population II
--- supernovae: general
%--- Galaxy: abundances
--- Galaxy: evolution
--- Galaxy: halo
}

\section{Introduction}

Recent spectroscopic studies have demonstrated the existence of numerous
carbon-enhanced, metal-poor (CEMP) stars that exhibit strong
enhancements of their neutron-capture elements, in particular at
metallicities [Fe/H]$ < -2.0$. This is believed to be due to mass
transfer in binary systems from former asymptotic giant-branch (AGB)
companions that underwent $s$-process nucleosynthesis during their
lifetimes \citep{McCl90}. However, a significant fraction of these stars
appear to exhibit large deviations from the scaled solar $s$-process
distribution of elemental abundance ratios, especially with regard to
their enhanced Eu \citep[e.g.,][]{Siva04}.

This observed discrepancy in Eu for some CEMP stars prompted
\citet{Hill00}, \citet{Cohe03}, \cite{Nomo04}, and \citet{Barb05} to
suggest that the large excess of Eu in these stars might be due to an
additional contribution from the $r$-process. \citet{BC05}, in their
suggested taxonomy of CEMP stars, refer to this class as
CEMP-$r/s$. \citet{Qian03} proposed an accretion-induced collapse
(AIC) of a white dwarf into a neutron star in a binary system as one
astrophysical scenario for enrichment of the surviving companion with
both $r$-process and $s$-process elements. As an alternative,
\citet{Zijl04} suggested that these double enhancements might be due to
the explosions of degenerate cores in AGB stars \citep[``type 1.5
supernova''; ] []{Iben83}. Previously, \citet{Nomo04} and
\citet{Barb05} suggested massive AGB stars ($\approx 8-10
M_\odot$)\footnote{The suggested mass range by \citet{Barb05}, $\approx
10-12 M_\odot$, is likely to be too high for stars that undergo
AGB evolution. The low metallicity of the CEMP-$r/s$ stars
pushes the appropriate mass range to even lower values \citep[$\approx 7-9 M_\odot$,
e.g.,][]{Umed99}.} to be the origin of these double enhancements.

In this paper we explore the astrophysical scenario suggested by
\citet{Nomo04} and \citet{Barb05}.  Specifically, we consider the
available constraints on a model in which $8-10 M_\odot$ stars in wide
binary systems may be invoked to explain the double enhancements of
$r$-process and $s$-process elements that results in the creation of
CEMP-$r/s$ stars. In \S~2 the observed abundances of CEMP-$r/s$ stars
are compared to the abundances from a low-metallicity AGB model
\citep{Gori00}, which implies additional contributions of $r$-processed
material to these stars. In \S~3, the wind pollution model is examined
to explain the double enhancements with $r$-process and $s$-process
elements in these stars. The efficiency of wind pollution by a supernova
required to be consistent with the Galactic chemical evolution of
$r$-process elements is then discussed. A brief summary of our
conclusions and a discussion of future areas for theoretical and
observational investigation of the CEMP-$r/s$ phenomenon is presented in
\S 4.

\section{CEMP-{\boldmath $r/s$} Stars}

Table~1 lists seven CEMP-$r/s$ stars with large enhancements of
neutron-capture elements reported in the recent literature
\citep{Aoki02, Cohe03, Barb05, Ivan05}, all of which were further
selected to have the lowest observed [Ba/Eu]\footnote{$[{\rm A}/{\rm B}]
= \log (N_{\rm A}/N_{\rm B}) - \log (N_{\rm A}/N_{\rm B})_\odot$} ratios
($< 0.4$). Note that the highly $r$-process-enhanced star CS~22892-052
\citep{Sned03a}, which marginally qualifies as a CEMP star, is added on
the last line of Table~1 for comparison purposes. With the exception of
this star, which exhibits a nearly solar $r$-process ratio of [Ba/Eu],
all of the CEMP-$r/s$ stars listed in Table~1 lie between the solar
$r$-process and $s$-process values of [Ba/Eu] \citep[$-0.69$ and +1.15,
respectively,][]{Arla99}.

These stars have similar metallicities ($-2.9 \le [{\rm Fe}/{\rm H}] \le -
2.3$) and excesses in their C and Ba abundances ($[{\rm C}/{\rm Fe}]
\sim [{\rm Ba}/{\rm Fe}] \sim +2$). These stars also exhibit very high
Pb abundances ($[{\rm Pb}/{\rm Fe}] \sim +3$), and hence belong to the
class of ``lead stars'' \citep{Aoki02, VanE03}. The over-production of
Pb is believed to be a consequence of the operation of an $s$-process
with a high neutron-to-seed ratio in an AGB star, owing to its low
metallicity \citep{Gall98, Gori00}. The similarity of the
abundance patterns amongst these stars implies that all the CEMP-$r/s$ stars
may have been formed in similar astrophysical environments.

\subsection{Comparison with a Low-Metallicity AGB Model}

Figure~1 compares the observed abundances of two representative stars from Table
1, HE~2148-1247 and CS~29497-030, to the abundances of elements predicted to be
found in the dredge-up material of a low-metallicity AGB model taken from
\citet{Gori00}, as shown by the thin-solid line. The metallicity of this model,
$[{\rm Fe}/{\rm H}] = - 1.3$, is clearly not as low as to be completely relevant
for the stars presented here. Nevertheless, we employ this comparison, since a
zero-metal AGB model by \citet{Gori01} appears to show a similar abundance
trend. The \citet{Gori00} result predicts a slightly lower [Ba/Eu] ratio
($+1.08$) than the solar $s$-process value ($+1.15$), but still significantly
higher than applies to the measured stellar abundances ($< +0.4$) listed in
Table~1. As can be seen in Figure~1, the abundance curve from the
low-metallicity AGB model (scaled to match the Ba abundances) does not appear to
account for any of the stellar abundances of Eu, Gd, and Dy (as well as Ho, Er, Yb, and
Hf for CS~29497-030), although the highly enhanced Pb (and Bi for CS~29497-030)
can be reasonably explained.

It should be noted that the [Ba/Eu] ratio in the atmosphere of the
observed stars could be expected to be lower than the dredge-up value
taken here, after this material is mixed with the envelopes of the
primary (former AGB) and secondary (post-accretion) stars. For example,
these envelopes might already have contained $r$-process matter at the
time of their formation \citep[e.g., from supernova-induced star
formation with production of $r$-process nuclei;][]{Ishi99}. It is
unlikely, however, that the mixture of envelope material will result in
such low [Ba/Eu] values ($< +0.4$) for all the stars considered here. In
fact, the CEMP-$r/s$ stars account for about $30 \%$ of all the CEMP
stars with the enhancements of $s$-process elements
\citep[e.g.,][]{Siva04}. On the other hand, non-CEMP-$r/s$ stars with a
high [Eu/Fe] ratio, relevant to those considered here (Table~1, $+1.6
\le {\rm [Eu/Fe]} \le +2.0$) are extremely rare -- CS~22892-052
\citep[${\rm [Eu/Fe]} = +1.64$,][]{Sned03a} and CS~31082-001
\citep[${\rm [Eu/Fe]} = +1.63$,][]{Hill02} are the only two such stars
with published high-quality abundance analyses. Such
stars account for only a few percent of the stars near $[{\rm Fe}/{\rm
H}] = - 3.0$ \citep{Bark05}.

%The average [Eu/Fe] for stars between $[{\rm Fe}/{\rm H}]
%= -3.0$ and $-2.0$ is about +0.5 \citep{Ishi04}, more than a factor of
%ten smaller than the values listed in Table~1.

\subsection{Are There Additional {\boldmath $r$}-Process Contributions?}

There exists the possibility that, under conditions of extremely high
neutron density ($\sim 10^{12} {\rm cm}^{-3}$), with a sufficiently
large exposure in an AGB star, one might obtain a low [Ba/Eu] value by
an ``$sr$-process'' \citep[e.g.,][]{Gori00}, a model that must be
investigated more thoroughly in the future. With currently available
data, however, the possibility of $r$-process contributions cannot be
excluded. As an exercise, the abundances of HE~2148-1247 and
CS~29497-030 are further compared with a simple mixture of the
abundances of the low-metallicity AGB model and the solar $r$-process
abundances \citep{Kapp89}, which is normalized to match the [Ba/Eu]
ratio in the star. This result is shown in Figure~1 by a thick-solid
line, together with the solar $r$-process (thick-dotted) and $s$-process
(thin-dotted) abundance curves, scaled to match the Ba abundance. Good
agreement can be seen for both stars, including the boosted Pb
abundances.

The non-detection of Th in these stars \citep[see also][]{John04} may
not be crucial, since theoretical studies show that the Th abundance can
be significantly lower than the scaled solar $r$-process curve, even
when good agreement is observed up to the third $r$-process-peak
elements \citep{Wana02, Wana03}. In addition, the upper limit on Th for
HE~2148-1247 (Figure~1a) still does not conflict with the mixture of the
solar $r$-process and $s$-process (assuming the Th abundance at the
birth of the star may be $\sim 0.3$~dex higher). Nevertheless, future
detection of Th (which cannot be synthesized by the $s$-process) would
strongly support a contribution from the $r$-process, although it
presents an observational challenge because of severe blending with CH lines
\citep{Nor97, Cohe03, John04}. Alternatively, measurements (or solid upper
limits) on the abundances of elements near the third $r$-process peak (Os, Ir,
and Pt), which are also not produced by the $s$-process, would be of particular
importance to support a contribution from the $r$-process (see Figure~1b).

Another way to check for possible contamination by the $r$-process may
be the (accurate) determination of isotopic ratios of, e.g., Eu. For
instance, the ratio $^{151}$Eu/($^{151}$Eu + $^{153}$Eu) would be
$\approx 0.5$ if there were a substantial contribution from the
$r$-process \citep{Sned02, Wana02, Aoki03a}. In contrast, the ratio
would be $\sim 0.6$ if the $s$-process dominated, as is found for some
of the CEMP stars with large enhancements of $s$-process elements, such
as LP~625-44 and CS~31062-050 \citep{Aoki03b}. Future accurate
measurements of the isotopic ratio of Eu (or other elements, if
possible) for the stars listed in Table~1 would be of special importance
to test for $r$-process contributions to their abundances.

\section{Double Enhancements by {\boldmath $8-10 M_\odot$} Stars}

The presence of $s$-process elements, along with large enhancements of
carbon ($[{\rm C}/{\rm Fe}] \sim +2$, Table~1) suggests that a
mass-transfer episode from a former AGB companion in a binary system
took place \citep{McCl90}. Thus, one major goal is to find an
astrophysical scenario, associated with an AGB star in a binary system,
in which the $r$-process might also occur. Recent nucleosynthesis
studies suggest that core-collapse supernovae, which include ``neutrino
winds'' \citep{Woos94, Wana01} and ``prompt explosions'' \citep{Sumi01,
Wana03} may be responsible for the production of the $r$-process
elements. It should be emphasized that all of these models suffer from
severe problems that remain to be solved \citep[e.g.,][]{Wana01, Wana03,
Jank05}, and no consensus has yet been achieved. Nevertheless,
remarkable agreements of the neutron-capture elements in some extremely
metal-poor stars, e.g., CS~22892-052 \citep{Sned03a} and CS~31082-001
\citep{Hill02}, with the scaled solar $r$-process curve strongly support
the idea that $r$-process elements originate from short-lived, massive
stars.

\subsection{Scenarios for Double Enhancements}

\citet{Qian03} suggested that the double enhancement in the CEMP-$r/s$
star HE~2148-1247 is due to the $s$-process occurring in an AGB star
member of a binary system, followed by the $r$-process taking place in a
subsequent AIC of the white dwarf remnant of the former AGB. The
nucleosynthetic outcome from an AIC event, {\it if it occurs}, can be
similar to that arising from a core-collapse supernova, although the
absence of an outer envelope in the former case may cause some
differences. The rate of the occurrence of the AIC process in the Galaxy
is highly uncertain, and perhaps no more than $\sim 10^{-4}$~yr$^{-1}$
\citep[e.g.,][]{Bail90} \citep[but see][]{Qian03}. This rarity seems to
be in conflict with the substantial fraction of CEMP-$r/s$ stars ($\sim
30\%$) among all the CEMP-$s$ stars currently observed. In addition, this
scenario involves {\it three} separate mass-transfer episodes --
transfer of $s$-process elements from a former AGB companion, mass
accretion onto the white dwarf remnant, and subsequent pollution of the
presently observed member of the system by $r$-process elements formed
during an AIC event of the white dwarf.  This may make such an event
extremely rare, although the possibility cannot be
excluded. Furthermore, an AIC is thought to only occur in close binary
systems, which is in conflict with the long periods observed for some of
the stars listed in Table~1 (see below). Note, however, that it remains
possible that the explosion may change the orbital period of the binary,
or even fractionate the pair into single stars.

Compared to the above model, the scenario suggested by \citet{Zijl04}
has the advantage that it involves only two mass-transfer steps --
transfer of $s$-process elements from an AGB companion followed by
pollution with $r$-process elements by a ``Type~1.5'' supernova
event. The nucleosynthetic outcome from a ``Type~1.5'' supernova may be
very similar to that of a Type~Ia supernova, in which
$r$-processing is not expected to be significant. \citet{Nomo04} and
\citet{Barb05} suggested an alternative scenario, in which the double
enhancements are due to a massive AGB star that may eventually collapse
to be an electron-capture supernova rather than a ``Type~1.5''
supernova.

Below we further examine the possibilities suggested by \citet{Nomo04} and
\citet{Barb05}, by investigation of a scenario in
which an $8-10 M_\odot$ star with a low-mass companion ($\sim 0.8
M_\odot$) in a wide binary system is responsible for the double
enhancements of $r$-process and $s$-process elements, resulting in
CEMP-$r/s$ stars. A star in this mass range is likely to undergo
$s$-processing during its AGB phase \citep{Nomo84}, although the amount
of the $s$-processed material produced and its expected abundance
distribution is uncertain. Subsequently, the degenerate O-Ne-Mg core of
this star may collapse by electron capture and explode \citep{Nomo84,
Nomo87, Hill84, Nomo88, Hash93, Jank05}; in such a scenario the $r$-process is
expected to take place \citep{Wana03}. This model also involves only
two mass-transfer episodes, as in \citet{Zijl04}.

The possibility of $s$-processing occurring in a $10 M_\odot$ star with
an O-Ne-Mg core has been suggested recently by \citet{Rito96} (see also
N. Iwamoto et al. 2005, in preparation, for a similar result with a $9
M_\odot$ star). These authors demonstrate that the reaction $^{22}{\rm Ne} (\alpha, n)
^{25}{\rm Mg}$ is efficient in such high-mass stars, owing to the high
temperature ($\gtrsim 3 \times 10^8$~K) reached at the base of the He convective
shell, suggesting that the likely occurrence of the $s$-process for stars in this
mass range. It is clear that a more quantitative study of the
$s$-process nucleosynthesis in this mass range is needed in the future, but the
results to date are certainly encouraging.

\subsection{Fate of an {\boldmath $8-10 M_\odot$} Star}

Nomoto (1984) pointed out that the final fate of $8 - 10 M_\odot$ stars
could be divided into the following two cases, depending on the highly
uncertain mass-loss rate.  (i) For stars in the mass range from 8
$M_\odot$ to $M_{\rm up}$, mass loss results in the ejection of the entire
envelope before the core mass reaches the Chandrasekhar limit, thereby
leaving an O-Ne-Mg white dwarf.  (ii) Stars in the mass range from
$M_{\rm up}$ to $10 M_\odot$ undergo electron-capture supernovae.  Here
$M_{\rm up}$ denotes the upper bound mass of the white dwarf
progenitors; if only the C+O white dwarfs are considered, $M_{\rm up}
\sim 8 M_\odot$ \citep[see, e.g.,][for the metallicity dependence of $M_{\rm
up}$]{Umed99}, while $8 M_\odot \le M_{\rm up} \le 10 M_\odot$ applies
to the progenitors of O-Ne-Mg white dwarfs \citep[see also the more recent
studies of][]{Rito96, Iben97, Rito99, Eldr04}.

Stars in the mass range $8-10 M_\odot$ that are found in close binary
systems, on the other hand, become helium stars that expand to red
giants. Subsequently, their helium envelope is lost by Roche-lobe
overflow and O-Ne-Mg dwarfs are formed \citep{Nomo84, Habe86}. Hence,
these stars do not undergo electron-capture supernovae \citep[but they
can undergo AICs;][] {Nomo91}\footnote{\cite{Pods04} concluded that such
stars in close binary systems result in electron-capture supernova
because the helium core mass is larger than the Chandrasekhar mass.
However, these stars would eventually leave O-Ne-Mg white dwarfs, when
considering the later evolutionary phases as described here. We thus
consider a wide binary system in this study.}.

\subsection{The Binarity of CEMP-{\boldmath $r/s$} Stars}

Currently, HE~2148-1247, CS~22948-027, CS~29497-034, CS~29526-110, and
CS~29497-030 (Table~1) have been found to be radial-velocity variables,
indicating their binarity. For CS~22948-027, CS~29497-034 \citep{Barb05},
and CS~29497-030 \citep{Sned03b}, the orbital periods are estimated to
be 426.5, 4130, and 342 days, respectively. These stars may belong to
the class of CH-star binaries with orbital periods of $\sim 1-10$ years,
as found by \citet{McCl90}, although no clear evidence of binarity for the
other two CEMP-$r/s$ stars in Table~1 has been obtained to date.

This high binary frequency \citep[see also][]{Luca05} implies that, {\it
if the current scenario is correct}, an electron-capture supernova
must only rarely, if ever, fractionate the pair into single stars. This is in contrast to
the neutrino-driven supernova from a more massive progenitor that may
obtain a large recoil velocity \citep[$\sim
500$~km~s$^{-1}$;][]{Sche04}. The lack of fractionation might result if the shock
arising from an O-Ne-Mg core is lifted too early after bounce \citep[$\sim
80$~ms,][]{Jank05} to obtain a large recoil by convective instability
\citep[$\sim 1$~s,][]{Sche04}, owing to the steep density gradient of
the outer edge of the core \citep{Nomo84}.

\subsection{Wind Pollution from an AGB Star}

For wide binary systems mass transfer operates through stellar winds,
rather than by Roche-Lobe overflow \citep{Boff88}. \citet{Theu96} estimated
the fraction of the mass captured by its companion ($1.5 M_\odot$) to be
$\sim 1-2$~\% of the mass lost by the wind (15~km~s$^{-1}$) from the AGB
star ($3 M_\odot$) with a period of 895~days and an orbital velocity of
36~km~s$^{-1}$, using a three-dimensional, smoothed particle
hydrodynamic simulation.

Note that most of the CEMP-$r/s$ stars under consideration here (five stars in
Table~1) have relatively high effective temperatures
\citep[$6000-7000$~K,][]{Aoki02, Cohe03, Ivan05}, which suggests that
these stars are main-sequence turn-off stars. For main-sequence
stars of $\approx 0.8 M_\odot$ with [Fe/H] $< -2$, the mass of the
convective envelope, $M_c$, is smaller by a factor of $\sim 10$ than that
for a stars with solar metallicity \citep[a few $10^{-3} M_\odot$, e.g., see
Table~1 in][]{Yosh81}. As a result, the dilution of the accreted
material is relatively small. Here we assume $M_c = 2 \times 10^{-3}
M_\odot$, which is one order of magnitude smaller than that of the Sun
\citep[$\sim 0.02 M_\odot$,][]{Bahc95}. If we take [C/Fe] = [Ba/Fe] = +2
and [Fe/H] = $-2.5$ as representative of the CEMP-$r/s$ stars in Table~1,
the accreted masses of C and Ba from the AGB onto its companion are
estimated to be $\sim 2 \times 10^{-5} M_\odot$ and $\sim 1 \times
10^{-10} M_\odot$, respectively.

Assuming the mass accretion rate to be $1\%$, according to
\citet[][although the binary system in their simulation is not
completely relevant to this study]{Theu96}, the masses of C and Ba
ejected from the AGB are estimated to be $\sim 2 \times 10^{-4} M_\odot$
and $\sim 1 \times 10^{-9} M_\odot$. The former is in good agreement
with the result from the stellar evolution calculation of a $9 M_\odot$
star by N. Iwamoto et al. (2005, in preparation). For the abundance of
Ba, a future study of the $s$-process nucleosynthesis in an $8-10
M_\odot$ model is needed to confirm the current hypothesis. Note that the
amount of C added by the subsequent supernova wind is negligible, owing
to its far less efficient accretion onto the secondary compared to that
from an AGB star (\S~3.6).

\subsection{Wind Pollution from a Supernova}

There is no currently available hydrodynamic study of the efficiency of
wind pollution by a supernova in a binary system.  Hence, we now
estimate the fraction of the ejected mass from a subsequent supernova
explosion which is captured by the companion star simply to be
$(R_2/2a_0)^2 f$, as in \citet{Pods02}. Here, $(R_2/2a_0)^2$ is the
geometric fraction of the companion, $R_2$ is the radius of the
secondary, $a_0$ is the initial orbital separation, and $f$ is the
enhancement (or reduction) factor.  The value of $f$ can be larger than
unity as the result of ``gravitational focusing'', if the ejecta
velocity is decelerated below the escape velocity from the surface of
the secondary. On the other hand, $f$ can be smaller than unity if the
ejecta velocity is large enough to strip the surface material from the
secondary.

Assuming the masses of the primary and secondary to be $9 M_\odot$ and
$1 M_\odot$, respectively, with $R_2 = 1 R_\odot$ and an initial orbital
period of one year, we obtain $(R_2/2a_0)^2 \sim 1 \times 10^{-6}$
(which reduces to $\sim 1 \times 10^{-7}$, if we change the initial
orbital period to five years). Note that the orbital period would become
as twice as large as its initial value after the explosion of the
primary, owing to the reduction of its mass to $\sim 1.3 M_\odot$, even
if the orbital separation were unchanged. The accreted material is
further diluted with the mass of the convective zone of the secondary,
e.g., $M_c = 2 \times 10^{-3} M_\odot$ (see \S~3.4). This results in the
required mass of Eu produced per supernova event (from a star of initial
mass $9 M_\odot$) is $M_{\rm Eu} \sim 3 \times 10^{-7} f^{-1} M_\odot$,
to obtain $[{\rm Eu}/{\rm Fe}] \sim +2$ for the secondary star with
$[{\rm Fe}/{\rm H}] \sim - 2.5$.

\subsection{Consistency with Galactic Chemical Evolution}

A Galactic chemical evolution study shows that the required mass of Eu
per supernova event in order to account for its solar value is $\sim 1
\times 10^{-7} M_\odot$ \citep[e.g.,][]{Wana05}, if {\it all}
core-collapse supernovae equally contribute to its
enrichment. \citet{Ishi99} have suggested, however, that the supernova
progenitors that contribute to the chemical evolution of Eu (as
representative of $r$-process elements) must be limited to a small range
\citep[$\sim 10\%$ of all supernova events, e.g., $8-10 M_\odot$ or
$20-25 M_\odot$, see also][]{Tsuj00, Ishi04, Wana05}. This leads to a
natural explanation of the large star-to-star scatter of the $r$-process
elements (e.g., Eu) relative to iron (by more than two orders of
magnitude) that can be seen in extremely metal-poor halo stars.

If one assumes that the stars of $8-10 M_\odot$ (i.e., $M_{up} = 8 M_\odot$, see
\S~3.2) are the dominant source of Eu, the mass of Eu ejected per
explosive event estimated from a Galactic chemical evolution study
should be increased to $\sim 3 \times 10^{-7} M_\odot$ \citep{Ishi99,
Ishi04}. This follows because the mass range $8-10 M_\odot$
accounts for about $30\%$ of all supernova events, when assuming a
Salpeter initial mass function. The mass of Eu per event would be $\sim
1 \times 10^{-6} M_\odot$ if $M_{\rm up}$ were, e.g., $9.5 M_\odot$,
since the mass range of $9.5-10 M_\odot$ accounts for only about $10\%$
of all supernova events. Note that further restriction of the mass range
(e.g., $M_{up} = 9.9 M_\odot$) would lead to larger star-to-star scatter
of [Eu/Fe] values than those observed in extremely metal-poor stars. It
should be also noted that the amount of Eu from a supernova event
estimated here seems reasonable from a nucleosynthetic point of view
\citep[e.g., from the neutrino wind scenario as a possible
explanation,][]{Wana01, Wana02}.

Thus, the estimate from the wind model (\S~3.5) and the constraint from
Galactic chemical evolution above are in good agreement when assuming $f
\sim 1$. However, an enhancement of the accretion by gravitational
focusing is needed ($f \sim 10$), if we assume an initial orbital period
of five years pertains.  This shows that wind pollution by a supernova
explosion is far less effective than that by an AGB star. In fact, the
efficiency of accretion for the AGB star is more than four orders of
magnitude larger than that estimated by the simple geometric fraction
(i.e., $f > 10^4$), owing to its small expansion velocity (comparable to
the orbital period of the system).

It should be noted that the contribution of $s$-processed material
(e.g., Ba) produced for stars of $8-10 M_\odot$ (which are short-lived
stars) must be a negligible contributor to the Galactic chemical
evolution of neutron-capture elements. This is required in order to be
consistent with observations of non-CEMP stars having [Ba/Eu] values
very close to the solar $r$-process ratio, with no sign of an increase
owing to the $s$-process for [Fe/H] $< -2.5$ \citep{John01}. The mass of
Ba per supernova event (from a star of initial mass $9 M_\odot$) due to
the $r$-process can be estimated to be $\sim 3 \times 10^{-6} M_\odot
f^{-1}$, assuming $M_{\rm Eu} \sim 3 \times 10^{-7} M_\odot f^{-1}$
applied above (\S~3.5) and the solar $r$-process ratio of Ba/Eu \citep[=
9.29,][]{Arla99}. On the other hand, the estimated mass of Ba from the
$s$-process during the AGB phase (for a star with an initial mass of $9
M_\odot$) is $\sim 1 \times 10^{-9} M_\odot$ (\S~3.4). This is
negligible compared to the $r$-process contribution considered here,
when assuming $f \sim 1-10$.

%being consistent with the spectroscopic results of
%the [Ba/Eu] values with mostly the solar $r$-process ratio.

%This results
%from the higher efficiency of the wind pollution by an AGB star (with
%smaller amounts of $s$-processed material).

\subsection{Enhancement of the Accretion Efficiency}

The discussion above demonstrates that the accretion efficiency onto the
secondary should not be reduced with respect to its geometric fraction,
leading to $f \gtrsim 1$. For supernova explosions with a typical
explosion energy ($\sim 10^{51}$~ergs), the velocity of the inner ejecta
is expected to be a few thousand km~s$^{-1}$ \cite[e.g.,][]{Shig94}. This is
larger than the escape velocity from the secondary (e.g,
$618$~km~s$^{-1}$ for the Sun), which may result in $f < 1$. However, a
collapsing O-Ne-Mg core is expected to lead to a neutrino-powered
explosion with a rather low explosion energy \citep[a few times
$10^{50}$~ergs;][]{Jank05}. Furthermore, its innermost ejecta, in which
the $r$-process is expected to operate, may expand rather slowly. In
fact, the core-collapse supernova from a more massive progenitor ($> 20
M_\odot$) is considered to suffer from fallback onto the remnant
\citep{Umed02, Umed03}, in which the expansion velocity of the inner
ejecta becomes zero at some point. If the innermost ejecta from a
collapsing O-Ne-Mg core expands slowly (e.g., $\lesssim$ a few
hundred~km~s$^{-1}$) without substantial fallback, the accretion can be
significantly enhanced by gravitational focusing. This may result in $f$
becoming larger than unity. It is obvious, however, that a detailed
hydrodynamic study will be needed in the future to estimate
quantitatively the efficiency of the wind-pollution model discussed
here.

\section{Conclusions}

The abundances of CEMP stars with large enhancements of $s$-process
elements, but with the lowest [Ba/Eu] ratios ($< +0.4$), disagree with
the predicted elemental abundance patterns from contemporary
low-metallicity AGB models, and seem to require an additional
$r$-process contribution. We have investigated a model in which these
CEMP-$r/s$ stars could be accounted for by an $8-10 M_\odot$ star in a
wide binary system that is responsible for enrichment with $s$-process
elements during its AGB phase, and with $r$-process elements by the
subsequent supernova explosion of its collapsing O-Ne-Mg core. It should
be cautioned, however, that the expected $s$-process signature resulting
from the AGB stage in stars in this mass range, as well as the
$r$-process abundance signature of the subsequent core collapse of stars
of this mass, are still not well known.

The estimated mass of Eu (as representative of $r$-process elements)
captured by the secondary, through the wind from the supernova, is in good
agreement with the constraint obtained from a Galactic chemical
evolution study, at least when the initial orbital separation is taken to be
$\sim 1$ year. However, the efficiency of wind pollution from the
supernova must be enhanced by a factor of $\sim 10$ when assuming an
initial orbital separation to be $\sim 5$ years. It is suggested that
the expansion velocity of the supernova's innermost ejecta, in which the $r$-process
has taken place, must be significantly slow, resulting in an enhancement of
accretion efficiency by gravitational focusing.

Future theoretical studies of $s$-process and $r$-process
nucleosynthesis in $8-10 M_\odot$ stars, as well as a full hydrodynamic
study of wind pollution during the supernova explosion in a wide binary
system, are needed before one can draw firm conclusions. Future
comprehensive spectroscopic studies of CEMP-$r/s$ stars, in particular
measurements of their Pt-peak (Os, Ir, and Pt) and Th abundances, and/or
isotopic Eu measurements, are also of special importance to confirm the
$r$-process contribution to these stars.

\acknowledgements

We would like to acknowledge W. Aoki, S. Goriely, and H. Umeda for
helpful discussions. We also acknowledge the contributions of an
anonymous referee, which led to clarification of a number of points in
our original manuscript. 

This work was supported in part by a Grant-in-Aid for the Japan-France Integrated
Action Program (SAKURA), awarded by the Japan Society for the Promotion of Science, and
Scientific Research (15204010, 16042201, 16540229, 17740108), and from the 21st
Century COE Program (QUEST) from the Ministry of Education, Culture, Sports,
Science, and Technology of Japan. T.C.B. acknowledges partial support from a
series of grants awarded by the US National Science Foundation, most recently,
AST 04-06784, as well as from grant PHY 02-16783; Physics Frontier Center/Joint
Institute for Nuclear Astrophysics (JINA).

%\newpage
%\clearpage

%\end{document}

\begin{deluxetable}{cllllllc}
%\footnotesize
\tablecaption{Abundance Ratios}
\tablewidth{0pt}
\tablehead{
\colhead{Star} &
\colhead{[Fe/H]} &
\colhead{[C/Fe]} &
\colhead{[Ba/Fe]} &
\colhead{[Eu/Fe]} &
\colhead{[Ba/Eu]} &
\colhead{[Pb/Fe]} &
\colhead{Reference}}
\startdata
HE~2148-1247 & $-$2.3  & +1.91 & +2.36 & +1.98 &   +0.38 & +3.12 & 1\\
CS~22948-027 & $-$2.47 & +2.43 & +2.26 & +1.88 &   +0.38 & +2.72 & 2\\
CS~29497-034 & $-$2.90 & +2.63 & +2.03 & +1.80 &   +0.23 & +2.95 & 2\\
CS~29526-110 & $-$2.38 & +2.2  & +2.11 & +1.73 &   +0.38 & +3.3  & 3\\
CS~22898-027 & $-$2.25 & +2.2  & +2.23 & +1.88 &   +0.35 & +2.84 & 3\\
CS~31062-012 & $-$2.55 & +2.1  & +1.98 & +1.62 &   +0.36 & +2.4  & 3\\
CS~29497-030 & $-$2.57 & +2.30 & +2.32 & +1.99 &   +0.33 & +3.65 & 4\\
CS~22892-052 & $-$3.10 & +0.95 & +0.99 & +1.64 & $-$0.65 & +1.20 & 5\\
\enddata

References.--- 1~\citet{Cohe03}; 2~\citet{Barb05}; 3~\citet{Aoki02};
4~\citet{Ivan05}; 5~\citet{Sned03a}

\end{deluxetable}

%\clearpage

\begin{figure}
\epsscale{2.}
\plotone{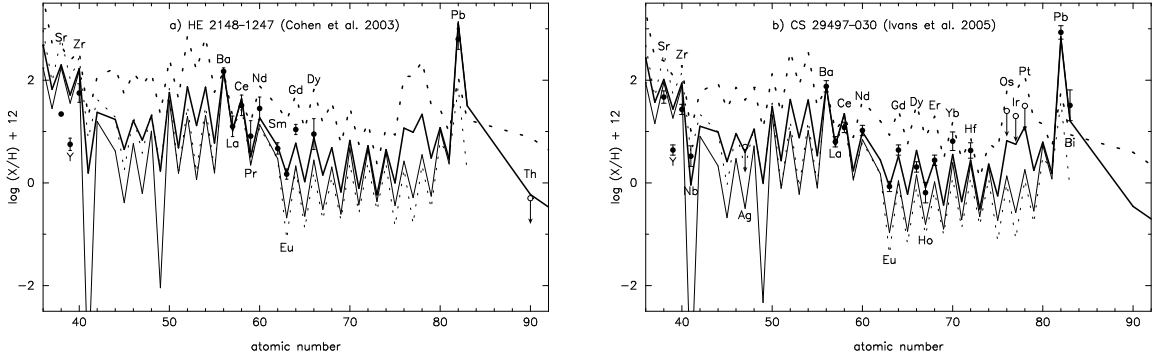}

\caption{Abundances in (a) HE~2148-1247 and (b) CS~29497-030, compared
with the solar $s$-process (thin-dotted line), $r$-process (thick-dotted
line), a low-metallicity AGB model \citep[thin-solid line,][]{Gori00},
and a mixture of the latter two (thick-solid line, see the text). The
abundances are vertically scaled to match the Ba abundance. For some
elements (Os, Ir, Pt, and Th), only an upper limit is shown (open
circle with down arrow).}

\end{figure}

\end{document}